\documentclass[twocolumn,aps,superscriptaddress,showpacs,nofootinbib,floatfix]{revtex4}
\usepackage{epsfig,bm,feynmf}
\usepackage{graphics}
\usepackage{amsmath}
\usepackage[normalem]{ulem}  
\usepackage[dvips]{color} 

\renewcommand\sout{\bgroup \color{red} \ULdepth=-.5ex \ULset}

\begin{document}


\title{Elliptic flows of light nuclei}


\author{Xuejiao Yin}\email{yinxuejiao@stu.scu.edu.cn}
\affiliation{Department of Physics, Sichuan University, Chengdu 610064, China}
\author{Che Ming Ko}\email{ko@comp.tamu.edu}
\affiliation{Cyclotron Institute and Department of Physics and Astronomy, Texas A$\&$M University, College Station, TX 77843, USA}
\author{Yifeng Sun}\email{sunyfphy@physics.tamu.edu}
\affiliation{Cyclotron Institute and Department of Physics and Astronomy, Texas A$\&$M University, College Station, TX 77843, USA}
\author{Lilin Zhu}\email{zhulilin@scu.edu.cn}
\affiliation{Department of Physics, Sichuan University, Chengdu 610064, China}



\begin{abstract}
Using the coalescence model based on nucleons from a blast-wave model with its parameters fitted to the measured proton transverse momentum spectrum and elliptic flow in heavy ion collisions at the Relativistic Heavy Ion Collider, we study the elliptic flows of light nuclei in these collisions. We find that to describe the measured elliptic flows of deuterons (anti-deuterons) and tritons (helium-3) requires that the emission source for nucleons of high transverse momentum is more elongated along the reaction plane than in the perpendicular direction.  Our results thus suggest that the elliptic flows of light nuclei can be used to study the nucleon emission source in relativistic heavy ion collisions.
\end{abstract}

\maketitle

\section{introduction}

There have been many studies in literatures on light nuclei production from heavy ion collisions at the Relativistic Heavy Ion Collider (RHIC) and the Large Hadron Collider (LHC). Most of these studies are concerned with their yields, and the experimental data from RHIC~\cite{Adler:2001uy,Abelev:2010rv,Agakishiev:2011ib} and LHC~\cite{Adam:2015yta,Adam:2015vda} can be successfully described by both the statistical model~\cite{Andronic:2010qu,Cleymans:2011pe} and the coalescence model~\cite{Chen:2012us,Chen:2013oba,Shah:2015oha,Sun:2015jta,Sun:2015ulc,Zhu:2015voa}. For the transverse momentum spectra of light nuclei, the coalescence model has also been shown to give a good description of the experimental data~\cite{Sun:2015jta,Sun:2015ulc}.

None of these theoretical studies has addressed the elliptic flows of light nuclei except Ref.~\cite{Oh:2009gx}, in which it is found that the deuteron elliptic flow obtained from the coalescence of freeze-out nucleons from a relativistic transport (ART) model~\cite{Li:1995pra,Li:2001xh} is similar to that produced via the reactions $NN\leftrightarrow\pi d$ in the transport model, and both give reasonable description of the experimental data from the PHENIX Collaboration at RHIC~\cite{Afanasiev:2007tv}.  In a recent study by the STAR Collaboration at RHIC~\cite{Adamczyk:2016gfs}, it has been shown that measured elliptic flows of light nuclei such as ${\rm d}$, $\bar{\rm d}$, ${\rm t}$, and $^3{\rm He}$ are consistent with the results calculated with the coalescence model that uses freeze-out nucleons from a multiphase transport (AMPT) model~\cite{Lin:2004en}.  On the other hand, the experimental data cannot be described by a blast-wave model with its parameters fitted to the proton transverse momentum spectrum and elliptic flow.  To understand the reasons for the failure of the blast-wave model and the success of the coalescence model in describing the elliptic flows of light nuclei, we investigate in the present study if the coalescence model carried out with the nucleon distributions from the blast-wave model can also describe the elliptic flows of light nuclei. We find that for this to be the case, high momentum nucleons need to be emitted more likely near the reaction plane than out of the reaction plane, compared to that for low momentum nucleons. Our results thus suggest that studying the elliptic flows of light nuclei provides the possibility of probing the nucleon emission source in relativistic heavy ion collisions, thus complementing the method that uses the Hanbury- Brown Twiss (HBT) interferometry of identical particles emitted at freeze-out~\cite{Bertsch:1988db,Pratt:1990zq,Mrowczynski:1992gc}.

The paper is organized as follows. In Section~\ref{blast}, we introduce the blast-wave model and describe how its parameters are determined.  The coalescence model is then described in Section~III. In Section~\ref{results}, we show results from our study on the transverse momentum spectra and elliptic flows of ${\rm d}$ ($\bar{\rm d}$) and ${\rm t}$ ($^3{\rm He}$) obtained from the blast-wave model without and with the space-momentum correlation.  Finally, a summary is given in Section~\ref{summary}.

\section{the blast wave model}\label{blast}

In the blast-wave model for particle production in relativistic heavy ion collisions, the Lorentz-invariant thermal distribution $f(x,p)$ of protons or neutrons at the kinetic freeze-out temperature $T_K$ is given by
\begin{equation}
f(x,p)=\frac{2\xi}{(2\pi)^3}\exp\{-p^\mu u_\mu/T_K^{}\},
\end{equation}
if the effect of quantum statistics is safely neglected at high temperature. In the above, $\xi$ is the fugacity of the proton or neutron, the four-momentum $p^\mu$ is
\begin{equation}\label{p}
p^\mu = (p^0,\bm{p}) = (m_T^{}\cosh y, p_T^{} \cos\phi_p,
p_T^{}\sin\phi_p, m_T^{} \sinh y),
\end{equation}
and the flow four-velocity $u_\mu(x)$ is
\begin{eqnarray}\label{u}
u^\mu &=& \cosh\rho ( \cosh\eta, \tanh\rho \cos\phi_b, \tanh\rho
\sin\phi_b, \sinh\eta).\nonumber \\
\end{eqnarray}
In Eqs.(\ref{p}) and (\ref{u}), $y$ is the energy-momentum rapidity, $m_T^{}$ ($= \sqrt{m^2 + p_T^2}$) is the transverse mass with $m$ and $p_T$ being the mass and transverse momentum of the nucleon, respectively, and $\eta$ and $\rho$ are the longitudinal and transverse flow rapidity $\eta=\frac{1}{2}\ln\frac{t+z}{t-z}$ and $\rho=\frac{1}{2}\ln\frac{1+|\bm\beta|}{1-|\bm\beta|}$, respectively, with $\bm\beta$ being the transverse flow velocity.  We have followed the usual convention to define the $z$-axis along the beam direction and the $x$-axis in the reaction plane of a heavy ion collision. The angles $\phi_p$ and $\phi_b$ are, respectively, the azimuthal angles of the nucleon momentum and the transverse flow velocity with respect to the $x$-axis.

From the nucleon momentum distribution, one can obtain the nucleon invariant momentum distribution as
\begin{eqnarray}
E\frac{d^3N}{d^3{\bf p}}=\frac{d^3N}{p_Tdp_Tdyd\phi_p}=\int_{\Sigma^\mu}d^3\sigma_\mu p^\mu f(x,p),
\end{eqnarray}
where $\Sigma^\mu$ is the freeze-out hyper-surface and $\sigma_\mu$ is its covariant normal vector.  In the case that $\Sigma^\mu$ depends only on the proper freeze-out time $\tau$, $\sigma_\mu$ is then given by
\begin{eqnarray}
d^3\sigma_\mu=(\cosh\eta,0,0,-\sinh\eta)\tau rdrd\eta d\phi,
\end{eqnarray}
with $\phi$ being the azimuthal angle of the nucleon transverse position vector.

Using the relations 
\begin{eqnarray}
p^\mu u_\mu&=&m_T\cosh\rho\cosh(\eta-y)-p_T\sinh\rho\cos(\phi_p-\phi_b),\nonumber\\
p^\mu d^3\sigma_\mu&=&\tau m_T\cosh(\eta-y)d\eta rdr d\phi,
\end{eqnarray}
and taking the particle emission time to be $\tau_0$, the invariant nucleon momentum spectrum can then be written as 
\begin{widetext}
\begin{eqnarray}
\frac{d^3N}{p_Tdp_Tdyd\phi_p}=\frac{2\xi\tau_0}{(2\pi)^3}\int_{\Sigma^\mu} d\eta rdr d\phi m_T\cosh(\eta-y)\exp{\left[-\frac{m_T\cosh\rho\cosh(\eta-y)-p_T\sinh\rho\cos(\phi_p-\phi_b)}{T_K}\right]}.
\end{eqnarray}
\end{widetext}

For simplicity, we assume that the longitudinal flow is boost invariant, so that the longitudinal flow rapidity is equal to the energy-momentum rapidity, i.e., $\eta=y$, and that the distribution in the energy-momentum rapidity is uniform in midrapidity. In this case, the invariant nucleon momentum spectrum becomes simply
\begin{eqnarray}
&&\frac{d^3N}{p_Tdp_Tdyd\phi_p}=\frac{2\xi\tau_0}{(2\pi)^3}\int_{\Sigma^\mu}rdr d\phi m_T\nonumber\\
&&\times\exp{\left[-\frac{m_T\cosh\rho-p_T\sinh\rho\cos(\phi_p-\phi_b)}{T_K}\right]}.
\end{eqnarray}

Because of non-vanishing elliptic flow $v_2^{}$ in non-central heavy ion collisions, which is defined as the average over all nucleons in all events,
\begin{equation}\label{elliptic}
v_2^{} =\left\langle \frac{p_x^2 - p_y^2}{p_x^2 + p_y^2} \right\rangle
\end{equation}
with $p_x$ and $p_y$ being, respectively, the projections of the nucleon transverse momentum along the $x$ and $y$ axes in the transverse plane, which is perpendicular to the reaction plane, the transverse flow velocity is anisotropic with respect to the azimuthal angle $\phi_b$. Following Ref.~\cite{Oh:2009gx}, we parameterize the transverse flow velocity as
\begin{equation}
\bm{\beta} = \beta(r) \left[ 1 + \varepsilon(p_T^{}) \cos(2\phi_b)
\right] \hat{\bm{n}},
\label{beta}
\end{equation}
where $\hat{\bm{n}}$ is the unit vector in the direction of their transverse flow velocity $\bm{\beta}$, which is taken to be normal to the surface of the system defined below. The $p_T^{}$-dependent coefficient $\varepsilon$ is introduced to model the saturation of the nucleon elliptic flow at large $p_T^{}$ as observed in experiments, and is parameterized as
\begin{equation}\label{epsilon}
\varepsilon(p_T^{}) = c_1^{} \exp(-p_T^{}/c_2^{}).
\end{equation}
We further parameterize the radial flow velocity as
\begin{equation}\label{beta0}
\beta(r) = \beta_0 \left( \frac{r}{R_0} \right),
\end{equation}
where $R_0$ is a parameter related to the transverse size of the nucleon distribution in space and $r$ is the distance of the nucleon from the origin of the coordinate system in the transverse plane.

For the spatial distribution of nucleons, it is assumed to be uniformly distributed inside a cylinder with its axis along the longitudinal direction and having an elliptic shape in the transverse plane, given by
\begin{equation}
\left( \frac{x}{A} \right)^2 + \left( \frac{y}{B} \right)^2\le1,
\label{ellipse1}
\end{equation}
with $A$ and $B$ related to the spatial elliptic anisotropy or eccentricity~\cite{Lin:2001zk}, $s_2^{} =\langle (x^2 - y^2)/(x^2 + y^2)\rangle$, by
\begin{equation}
A = R_0 (1 + s_2^{}), \qquad B = R_0 (1 - s_2^{}).
\end{equation}
The spatial region as given by Eq.~(\ref{ellipse1}) can also be rewritten as
\begin{equation}\label{radius}
r \le R_0 \left[ 1 + s_2^{} \cos(2\phi) \right]
\end{equation}
in terms of the azimuthal angle $\phi$ of the nucleon position vector ${\bf r}$ in the transverse plane. It can be shown that the angle $\phi$ is related to the azimuthal angle $\phi_b$ of transverse flow by $\tan\phi=(B/A)^2\tan\phi_b$~\cite{Sun:2015jta}.

\begin{figure}[h]
\centerline{
\includegraphics[width=10cm]{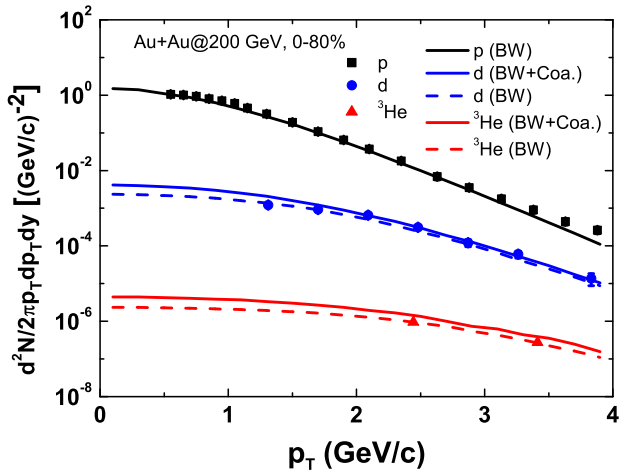}}
\caption{(Color online) Transverse momentum spectra of midrapidity proton, deuteron (anti-deuteron) and triton (helium-3) from the blast-wave model (dashed lines) and the coalescence model (solid lines) for Au+Au collisions at $\sqrt{s_{NN}}=200$ GeV and centrality of 0-80\%. Data are from Ref.~\cite{Abelev:2006jr} for proton, Ref.~\cite{Adler:2004uy} for deuteron (anti-deuteron), and Ref.~\cite{Abelev:2009ae} for triton (helium-3).}
\label{pt}
\end{figure}

\begin{figure}[h]
\centerline{
\includegraphics[width=10cm]{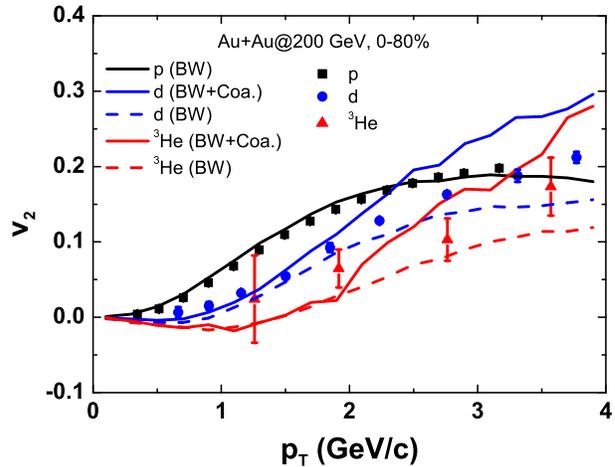}}
\caption{(Color online) Elliptic flows of midrapidity proton, deuteron (anti-deuteron) and triton (helium-3) from the blast-wave model (dashed lines) and the coalescence model (solid lines) for Au+Au collisions at $\sqrt{s_{NN}}=200$ GeV and centrality of 0-80\%. Data are from Ref.~\cite{Adamczyk:2015ukd} for proton, Ref.~\cite{Adamczyk:2016gfs} for deuteron (anti-deuteron) and triton (helium-3).}
\label{v2}
\end{figure}

For Au$+$Au collisions at center of mass energy $\sqrt{s_{NN}^{}} = 200$~GeV and centrality of 0-80\%, the parameters in the blast-wave model are taken as follows to reproduce the transverse momentum spectrum and elliptic flow of protons measured at RHIC. For the spatial distribution of nucleons, we use the parameters
\begin{eqnarray}
R_0 =10.0 \mbox{ fm},~s_2^{} = -0.04,~\tau_0^{}= 9.0 \mbox{ fm}/c,
\end{eqnarray}
For the transverse momentum distribution, we use the kinetic temperature $T_K^{} = 130$~MeV and
\begin{eqnarray}
\beta_0 = 0.67,~c_1^{} = 0.148,~c_2^{} = 2.12 \mbox{ GeV}/c
\label{eq:parameters}
\end{eqnarray}
for the flow velocity parameterized in Eqs.~(\ref{beta})-(\ref{beta0}). The measured total number of protons $N_p = 4.8$ is obtained from  the blast-wave model by using zero charge and baryon chemical potentials as well as a fugacity parameter $\xi=1.76$. These protons and a same number of neutrons are then uniformly distributed in the rapidity region $|y| \le 0.5$. The resulting transverse momentum spectrum and elliptic flow of protons are shown by the black solid lines in Figs.~\ref{pt} and \ref{v2}, respectively, and they are seen to reproduce very well the experimental data, shown by squares, from the STAR Collaboration~\cite{Abelev:2006jr,Adamczyk:2015ukd}.

\section{the coalescence model}\label{coalescence}

The coalescence model for nuclei production in heavy ion collisions is based on the sudden approximation of projecting out their wave functions from the wave functions of nucleons at freeze out.  As shown in Ref.~\cite{Mattiello:1996gq}, the number of nucleus of atomic number $A$ and consisting of $Z$ protons and $N$ neutrons produced in a heavy ion collision is given by the overlap integral of the Wigner function $f_A({\bf x}_1^\prime, ... ,{\bf x}_Z^\prime,{\bf x}_1^\prime, ... ,{\bf x}_N^\prime; {\bf p}_1^\prime, ... ,{\bf p}_Z^\prime,{\bf p}_1^\prime, ... ,{\bf p}_N^\prime,t^\prime)$ of the produced nucleus with the phase-space distribution function 
$f_p({\bf x},{\bf p},t)$ of protons and $f_n({\bf x},{\bf p},t)$ of neutrons at freeze out, which is normalized to the total proton number $N_p$ and neutron number $N_n$ according to $\int p^\mu d^3\sigma_{\mu}\frac{d^3{\bf p}}{E}f_p({\bf x}, {\bf p},t)=N_p$ and $\int p^\mu d^3\sigma_{\mu}\frac{d^3{\bf p}}{E}f_n({\bf x}, {\bf p},t)=N_n$, that is
\begin{eqnarray}
\label{coal}
&&\frac{dN_A}{d^3 {\mathbf P}_A}=g_A\int \Pi_{i=1}^Zp_i^\mu d^3\sigma_{i\mu}\frac{d^3{\bf p}_i}{E_i}f_p({\bf x}_i, {\bf p}_i,t_i)\nonumber\\
&&\times\int \Pi_{j=1}^Np_j^\mu d^3\sigma_{j\mu}\frac{d^3{\bf p}_j}{E_j}f_n({\bf x}_j, {\bf p}_j,t_j)\nonumber\\
&&\times f_A({\bf x}_1^\prime, ... ,{\bf x}_Z^\prime,{\bf x}_1^\prime, ... ,{\bf x}_N^\prime; {\bf p}_1^\prime, ... ,{\bf p}_Z^\prime,{\bf p}_1^\prime, ... ,{\bf p}_N^\prime;t^\prime)\nonumber\\
&&\times \delta^{(3)}\left({\bf P}_A-\sum_{i=1}^Z{\bf p}_i-\sum_{j=1}^N{\bf p}_j\right),
\end{eqnarray}
where $g_A=(2J_A+1)/2^A$ is the statistical factor for $A$ nucleons of spin $1/2$ to form a nucleus of angular momentum $J_A$.  As in Refs.~\cite{Chen:2003ava,Oh:2009gx}, the coordinate ${\bf x}_i$ and momentum ${\bf p}_i$ are those of the $i$-th nucleon in the center of mass of all emitted particles in the blast wave. The corresponding coordinate ${\bf x}_i^\prime$ and momentum ${\bf p}_i^\prime$, which appear in the nuclear Wigner function, are obtained by first Lorentz transforming to the rest frame of produced nucleus and then letting this nucleon to propagate freely with a constant velocity, given by the ratio of  its momentum and energy in the rest frame of the nucleus, until the time $t^\prime$ when the last nucleon in the nucleus freezes out.

For the light nuclei we consider in this study, ${\rm d}$, $\bar{\rm d}$, ${\rm t}$, and $^3{\rm He}$, we approximate their wave functions by those of the ground state of a harmonic oscillator with the oscillator constant adjusted to fit the empirical charge radii of these nuclei. The Wigner function for ${\rm d}$ and similarly for $\bar{\rm d}$ is then~\cite{Song:2012cd}
\begin{eqnarray}
f_2(\boldsymbol\rho,{\bf p}_\rho)=8g_2\exp\left[-\frac{\boldsymbol\rho^2}{\sigma_\rho^2}-{\bf p}_\rho^2\sigma_\rho^2\right],
\label{two}
\end{eqnarray}
with
\begin{eqnarray}\label{rel}
\boldsymbol\rho=\frac{{\bf x}_1^\prime-{\bf x}_2^\prime}{\sqrt{2}},\quad{\bf p}_\rho=\frac{{\bf p}_1^\prime-{\bf p}_2^\prime}{\sqrt{2}},\nonumber\\
\end{eqnarray}
where we have used the same mass $m$ for proton and neutron.

Similarly, the Wigner function for ${\rm t}$ and $^3{\rm He}$ is~\cite{Song:2012cd}
\begin{eqnarray}
&&f_3(\boldsymbol\rho,\boldsymbol\lambda,{\bf p}_\rho,{\bf p}_\lambda)\nonumber\\
&&=8^2g_3\exp\left[-\frac{\boldsymbol\rho^2}{\sigma_\rho^2}-\frac{\boldsymbol\lambda^2}{\sigma_\lambda^2}-{\bf p}_\rho^2\sigma_\rho^2-{\bf p}_\lambda^2\sigma_\lambda^2\right],
\label{three}
\end{eqnarray}
where $\boldsymbol\rho$ and ${\bf p}_\rho$ are similarly defined as in Eq.(\ref{rel}), and
\begin{eqnarray}
{\boldsymbol\lambda}&=&\frac{{\bf x}_1^\prime+{\bf x}_2^\prime-2{\bf x}_3^\prime}{\sqrt{6}},\nonumber\\
{\bf p}_\lambda&=&\frac{{\bf p}_1^\prime+{\bf p}_2^\prime-2{\bf p}_3^\prime}{\sqrt{6}}.
\end{eqnarray}

The width parameter $\sigma_\rho$ in Eq.(\ref{two}) is related to the charge mean-square radius of deuteron and
also the oscillator frequency $\omega$ in the harmonic wave function via~\cite{Song:2012cd}
\begin{eqnarray}
\langle r_{\rm d}^2 \rangle=\frac{3}{4}\sigma_\rho^2=\frac{3}{4m\omega},
\end{eqnarray}
where the second line follows if we use the relation $\sigma_\rho=1/\sqrt{m\omega}$ in terms of the oscillator frequency $\omega$ in the harmonic wave function.

For the width parameter $\sigma_\lambda$ in Eq.(\ref{three}), it is again related to the oscillator frequency by $\sigma_\lambda=1/\sqrt{m\omega}$ and is thus equal to $\sigma_\rho$.  Both $\sigma_\rho$ and $\sigma_\lambda$ are then determined from the oscillator constant via the mean-square charge radius of ${\rm t}$ or $^3{\rm He}$, that is~\cite{Song:2012cd}
\begin{eqnarray}
\langle r_{\rm t}^2\rangle=\frac{1}{m\omega}, \quad\langle r_{\rm ^3He}^2\rangle=\frac{2}{m\omega}.
\end{eqnarray}

\begin{table}[h]
\caption{{\protect\small Statistical factor ($g$), root-mean-square charge radius ($R_{\rm ch}$), oscillator frequency ($\omega$), and width parameter ($\sigma_\rho$, $\sigma_\lambda$) for deuteron and anti-deuteron (${\rm d}$, $\bar{\rm d}$), triton(${\rm t}$), and helium-3 ($^3{\rm He}$). Radii are taken from Ref.~\cite{Angeli:2013}}.} \label{tab}
\smallskip
\begin{tabular}{c|cccccc}
\hline\hline
Nucleus & $g$ & $R_{\rm ch}$ (fm) & $\omega$ (sec$^{-1})$ & $\sigma_\rho, \sigma_\lambda$ (fm) \\
\hline
${\rm d}$,$\bar{\rm d}$ & 3/4 & 2.1421 & 0.1739 & 2.473 \\
${\rm t}$ & 1/4 & 1.7591 & 0.3438 & 1.759 \\
$^3{\rm He}$ & 1/4 & 1.9661 & 0.5504 & 1.390 \\
\hline\hline
\end{tabular}
\end{table}

The statistical factors and the values of the width parameters in the Wigner functions for ${\rm d}$, $\bar{\rm d}$, ${\rm t}$, and $^3{\rm He}$ as well as the empirical values of their charge radii and the resulting oscillator constants are given in Table \ref{tab}.

\section{results}\label{results}

In the present section, we show the transverse momentum spectra and elliptic flows for ${\rm d}$ ($\bar{\rm d}$) and ${\rm t}$ ($^3{\rm He}$) calculated from the coalescence model using the phase-space distributions of protons and neutrons from the blast-wave model.

\subsection{Transverse momentum spectra}

The transverse momentum spectrum of deuteron (anti-deuteron) in Au+Au collisions at $\sqrt{s_{NN}}=200$ GeV and centrality of $0-80\%$ is shown by the blue solid line in Fig.~\ref{pt}. It reproduces very well the data from the PHENIX Collaboration~\cite{Adler:2004uy} shown by circles. For comparison, we also show by the blue dashed line the deuteron transverse momentum spectrum from the blast wave model, which is obtained by replacing the proton mass by that of the deuteron and fitting the deuteron total number by taking the deuteron fugacity to be $\xi_d=2.1$. It is seen to be very similar to that from the coalescence model except slightly smaller at transverse momentum below about 2 GeV.  Also shown by the red solid line in Fig.~\ref{pt} is the transverse momentum spectrum of triton (helium-3) from the coalescence model, which is slightly larger than the experimental data, shown by triangles, from the STAR Collaboration~\cite{Abelev:2009ae}.   Results from the blast-wave model, shown by the red dashed line, for the transverse momentum spectrum of triton (helium-3) can describe, on the other hand, the experimental data very well with  the fugacity of triton (helium-3) taken to be $\xi_t=3.9$.

\subsection{Elliptic flows}

The azimuthal angle $\phi_T$ distribution of the transverse momentum ${\bf p}_T$ of nucleus $A$ produced in a heavy ion collision can be generally written as
\begin{eqnarray}
&&f_A(p_T,\phi_T,y)=\frac{N_A(p_T,y)}{2\pi}\nonumber\\
&&~~~\times\left\{1+2\sum_n v_n(p_T,y)\cos[n(\phi_T-\Psi_n)]\right\},
\end{eqnarray}
where $\Psi_n$ is the $n$th-order event plane angle, and $N_A(p_T,y)$ and $v_n(p_T,y)$ are the number of such nucleus in rapidity $y$ and their $n$th-order anisotropic flows, respectively. In the present study, we are only interested in the elliptic flow $v_2$.  Also, we take the event plane angle $\Psi_2=0$ because we generate many nucleons from the blast wave to reduce the statistical fluctuations due to the small number of nucleons in an event.  In this case, the elliptic flow can be simply calculated from Eq.(\ref{elliptic}).

In Fig.~\ref{v2}, we show by the blue solid line the deuteron $v_2$ from the coalescence model. Compared to the experimental data, shown by circles, from the STAR Collaboration~\cite{Adamczyk:2016gfs}, it overestimates the data above 2 GeV. On the other hand, the deuteron $v_2$ from the blast-wave model, shown by the blue dashed line, is smaller than the experimental data at all values of transverse momentum. The same is for the $v_2$ of triton (helium-3), shown by the red solid line, that it is smaller than the experimental data from the STAR Collaboration~\cite{Adamczyk:2016gfs}, shown by triangles, at small $p_T$ but larger at larger $p_T$, while that from the blast-wave model, shown by the red dashed line, again underestimates the data at all $p_T$.

\begin{figure}[h]
\centerline{
\includegraphics[width=10cm]{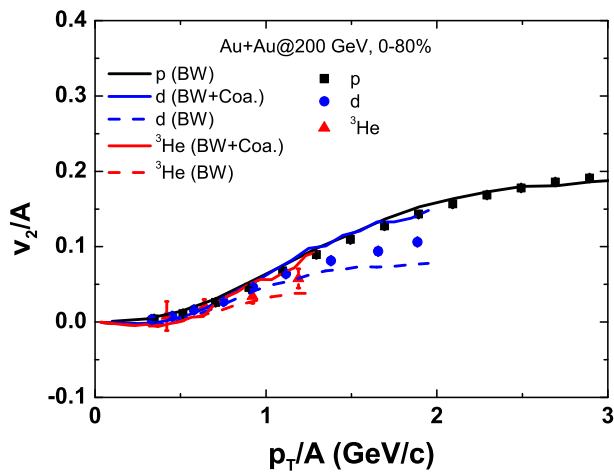}}
\caption{(Color online) Same as Fig.~\ref{v2} for constituent nucleon number scaled elliptic flows of midrapidity proton, deuteron (anti-deuteron) and triton (helium-3).}
\label{v2s}
\end{figure}

It is known that in the coalescence model for light nuclei production, the nucleon-number scaled elliptic flows  satisfy a scaling law that the elliptic flow of a nucleus per nucleon at transverse momentum $p_T$ is the same as a function of $p_T$ divided by the number of nucleons in the nucleus~\cite{Yan:2006bx,Zhu:2015voa}. For light nuclei considered here, $v_{2}(p_T/A)/A$ is then the same. This scaling would be exact if only nucleons of same momentum can coalesce to form a nucleus, corresponding to a width parameter in the Wigner function of the nucleus that is infinitely large~\cite{Chen:2006vc}.  For the elliptic flows of light nuclei from the present coalescence model, the nucleon number scaled elliptic flows of deuteron (anti-deuteron) and triton (helium-3) are shown in Fig.~\ref{v2s} by the blue and red solid lines, respectively, as functions of the scaled transverse momentum. It is seen that both follow closely the proton elliptic flow (black line) and thus indeed show a scaling behavior.  The scaling behavior of the elliptic flows of light nuclei is also seen in the experimental data at $p_T/A< 1$ GeV. At large $p_T/A$, the scaled elliptic flows of deuteron (anti-deuteron) and triton (helium-3) from the experiments are smaller than the measured proton elliptic flow. The corresponding results from the blast-wave model, shown by the blue and red dashed lines for deuteron (anti-deuteron) and triton (helium-3), respectively, are similar to the results from the coalescence model for $p_T/A<0.7$ GeV and thus also show a scaling behavior. At larger $p_T/A$, no scaling behavior is seen for the elliptic flows from the blast-wave model. Instead, we see a strong mass ordering that the scaled elliptic flow of deuteron (anti-deuteron) is smaller than the proton elliptic flow and that of triton (helium-3) is further smaller than that of deuteron (anti-deuteron).

\subsection{Effects of space-momentum correlation}

\begin{figure}[h]
\centerline{\includegraphics[width=10cm]{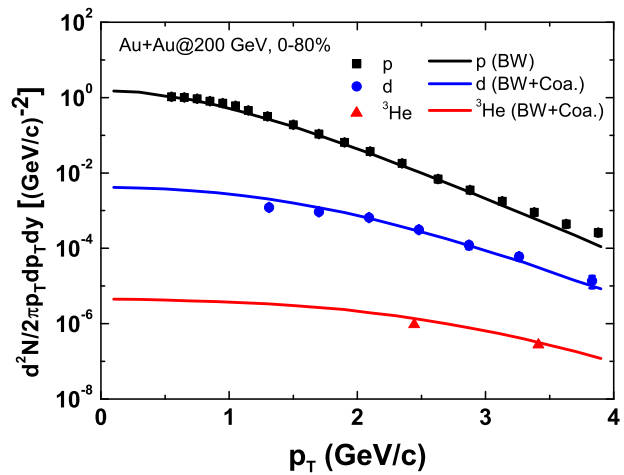}}
\caption{(Color online) Transverse momentum spectra of midrapidity proton, deuteron (anti-deuteron), and triton (helium-3) from the coalescence model based on a blast-wave model with  space-momentum correlation (solid lines) for Au+Au collisions at $\sqrt{s_{NN}}=200$ GeV at centrality of 0-80\%. Data are from Ref.~\cite{Abelev:2006jr} for proton, Ref.~\cite{Adler:2004uy} for deuteron (anti-deuteron), and Ref.~\cite{Abelev:2009ae} for triton (helium-3).}
\label{ptnew}
\end{figure}

Results shown in the previous subsection based on the coalescence model indicate that assuming the spatial distribution of nucleons is independent of their momenta fails to describe experimentally measured differential elliptic flows of light nuclei, particularly at high momenta where the theoretical results are larger than the experimental values.  Since light nuclei of large transverse momenta are produced in the coalescence model from nucleons of large transverse momenta, their elliptic flows can be reduced if these nucleons are more separated along the reaction plane than in the perpendicular direction. This is because such a space-momentum correlation reduces the production of light nuclei along the reaction plane and enhances it in the perpendicular direction.  Such a correlation can be introduced in the blast-wave model by requiring nucleons of $p_T> 0.9$ GeV to have a radius parameter $R_0=10~e^{0.23(p_T-0.9)}$ fm in Eq.(\ref{radius}) for $|p_{Tx}|>|p_{Ty}|$. For all other nucleons, the radius parameter is taken to be $R_0=10.0$ fm as in Section II.

\begin{figure}[h]
\centerline{\includegraphics[width=10cm]{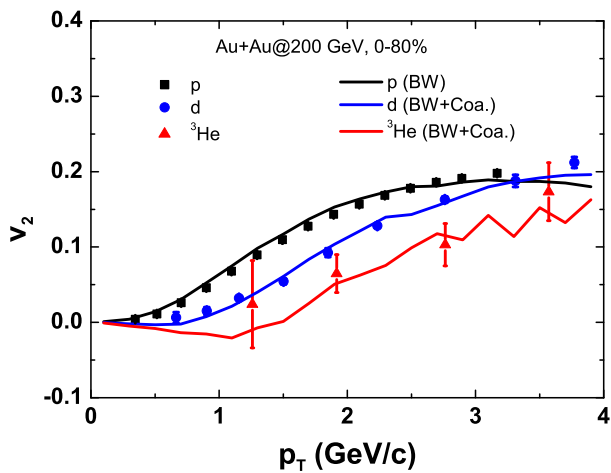}}
\caption{(Color online) Elliptic flows of midrapidity proton, deuteron (anti-deuteron), and triton (helium-3) at midrapidity from the coalescence model based on a blast-wave model with space-momentum correlation (solid lines) for Au+Au collisions at $\sqrt{s_{NN}}=200$ GeV at centrality of 0-80\%. Data are from Ref.~\cite{Adamczyk:2015ukd} for proton, Ref.~\cite{Adamczyk:2016gfs} for deuteron (anti-deuteron) and triton (helium-3).}
\label{v2new}
\end{figure}

As shown by black solid lines in Figs.~\ref{ptnew} and \ref{v2new}, the experimental proton transverse momentum spectrum and elliptic flow can still be very well reproduced with the same parameters given in Eqs.(16) and (17). For the transverse momentum spectra and elliptic flows of deuteron (anti-deuteron) and triton (helium-3) from the coalescence model based on the phase-space distribution functions of nucleons in the blast-wave model that includes the space-momentum correlation, they are shown, respectively, by blue and red solid lines in these figures. They are seen to describe very well the experimental data not only on the transverse momentum spectrum but also the elliptic flow of deuteron (anti-deuteron) and triton (helium-3).

\section{summary}\label{summary}

Using the coalescence model based on the nucleon phase-space distribution function from a blast-wave model, we have studied the transverse momentum spectra and elliptic flows of light nuclei in relativistic heavy ion collisions.  Assuming that the spatial distribution of nucleons in the system is independent of their momenta and fitting the parameters to the experimental measured proton transverse momentum spectrum and elliptic flow by the STAR collaboration at RHIC, we have obtained a good description of the measured transverse momentum spectra of deuteron (anti-deuteron) and triton (helium-3) but have failed to reproduce the measured elliptic flows of these nuclei, particularly at large transverse momenta.  We have attributed this failure of the coalescence model to the neglect of possible preference of nucleons of large transverse momenta to be more spread in space if their momenta are along the reaction plane than perpendicular to it.  Allowing a nucleon phase-space distribution that includes such space-momentum correlations in the blast-wave model indeed leads to a good description of the elliptic flows of deuteron (anti-deuteron) and triton (helium-3) as well.  Our study thus indicates that the elliptic flows of light nuclei are sensitive to the space and momentum correlations of nucleons at kinetic freeze out and are thus a possible probe of the nucleon emission source in relativistic heavy ion collisions.

\section*{Acknowledgements}

We thank Lie-Wen Chen, Zi-Wei Lin, Kaijia Sun, and Jun Xu for helpful communications and/or discussions.  One of the authors (C.M.K.) is grateful to the Physics Department at Sichuan University for the warm hospitality during his visit when this work was carried out.  This work was supported in part by the NSFC of China under Grant no. 11205106, the US Department of Energy under Contract No. DE-SC0015266, and The Welch Foundation under Grant No. A-1358.

\bibliography{references}

\end{document}